\documentclass{article}

\usepackage[preprint]{neurips_2026}

\usepackage[utf8]{inputenc} 
\usepackage[T1]{fontenc}    
\usepackage{hyperref}       
\usepackage{url}            
\usepackage{booktabs}       
\usepackage{amsfonts}       
\usepackage{nicefrac}       
\usepackage{microtype}      
\usepackage{xcolor}         

\usepackage{tabularx}
\usepackage{subcaption}
\usepackage{graphicx}
\usepackage{booktabs}
\usepackage{multirow}
\usepackage{arydshln}
\usepackage{amsmath}

\title{MADB: A Large-Scale Music Aesthetics Dataset with Professional and Multi-Dimensional Annotations}

\author{
    Sirui Zhang$^{1,2}$, Tianle Wang$^{1,2}$, Xinyi Tong$^{1,2}$, Peiyang Yu$^{1,2}$, Jishang Chen$^{1,2}$, \And
    Liangke Zhao$^{1,2}$, Haoxin Zhang$^{1,2}$, Duo Xu$^{2,3,*}$, Xin Jin$^{2,4,*}$, Feng Yu$^{1,*}$, Songchun Zhu$^{2,5,*}$\\[0.5em]
    $^{1}$Central Conservatory of Music, China\\
    $^{2}$Beijing Institute for General Artificial Intelligence\\
    $^{3}$Tianjin Conservatory of Music\\
    $^{4}$Beijing Electronic Science and Technology Institute\\
    $^{5}$Peking University\\[0.25em]
    $^{*}$Corresponding authors\\
}

\begin{document}

\maketitle

\begin{abstract}
Music aesthetic assessment is a challenging yet underexplored problem, requiring models to capture fine-grained, multi-dimensional human perceptual judgments.
Progress in this area has been limited by the lack of large-scale datasets with structured aesthetic annotations.

We introduce MADB, a large-scale dataset and benchmark comprising 9,999 tracks annotated by 30 trained annotators.
Each track is rated by around 10 annotators across 10 perceptual dimensions and one overall score, with additional textual comments for multimodal analysis.

We establish a unified evaluation framework over multiple pretrained models.
Results reveal substantial gaps between model predictions and human judgments, exposing key limitations of current approaches.

MADB provides a new benchmark for human-aligned music understanding.
Project page: https://github.com/knownree/madb
\end{abstract}

\begin{figure}[htbp]
    \centering
    \includegraphics[width=1\linewidth]{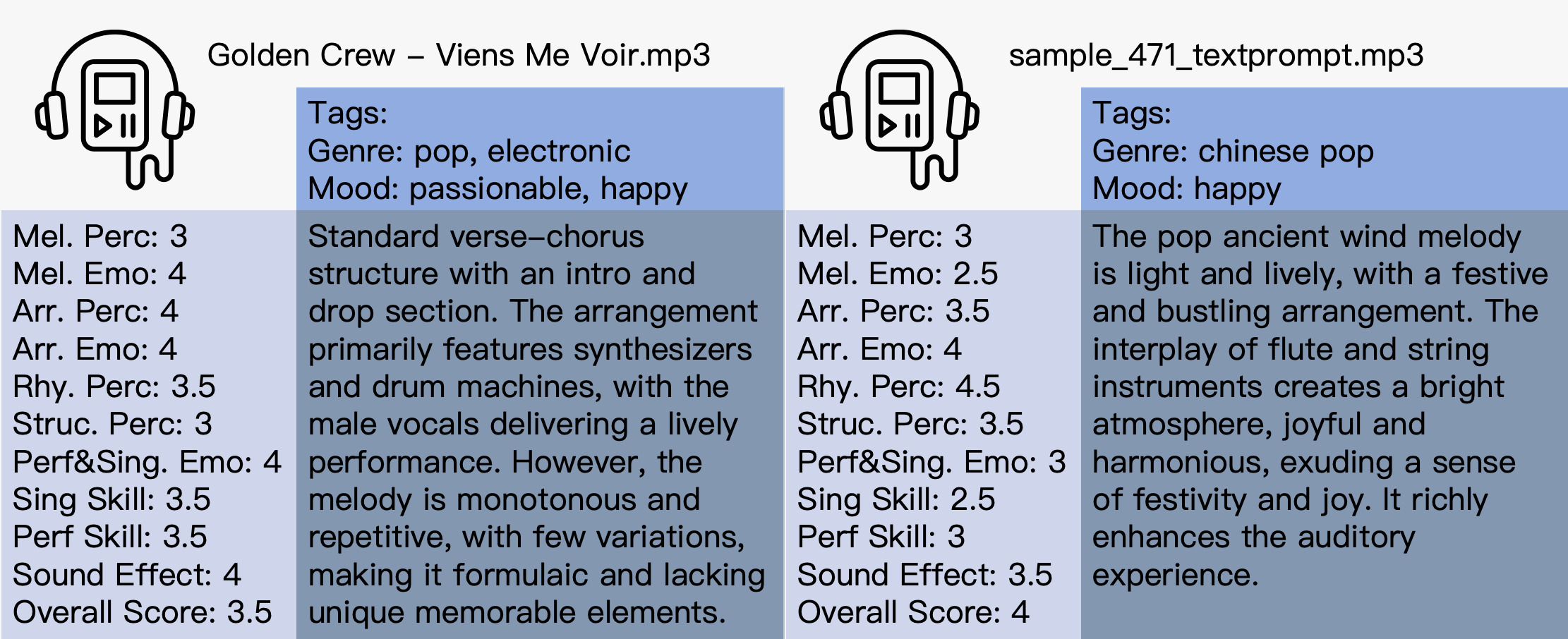}
    \caption{Annotation examples}
    \label{fig:anno}
\end{figure}

\section{Introduction}

The rapid development of generative music models~\cite{musiclm,LeVo, YUE} has led to a surge in AI-generated music, creating an urgent need for automatic evaluation aligned with human aesthetic preferences. 
The rapid expansion of AI-generated music not only creates diverse application scenarios for automatic aesthetic evaluation, but also elevates it to a critical component in modern music generation pipelines, with applications in content filtering, recommendation, and reinforcement learning from human feedback (RLHF).

Despite these advances, current models often fail to produce music that aligns with human expectations, exhibiting issues such as structural incoherence and limited expressiveness~\cite{musiclm, musicgen}. 
This gap highlights the need for reliable aesthetic evaluation models that can accurately reflect human perception.

Music aesthetic assessment is inherently challenging. 
It is subjective and exhibits substantial inter-individual variability, while also requiring the modeling of relatively stable and generalizable criteria grounded in domain knowledge. 
Moreover, aesthetic evaluation in music relies heavily on expertise: trained musicians can provide fine-grained and technically grounded assessments, explaining not only whether a piece is preferred but also why it is perceived as aesthetically effective or flawed. 
This dual requirement of subjectivity and expertise makes large-scale, high-quality annotation particularly challenging.

However, existing datasets for music aesthetic evaluation remain limited in scale, number of annotators, and coverage of perceptual dimensions, and often rely on single-score annotations without richer supervision such as multi-dimensional ratings or textual feedback. 

To address these challenges, we make the following contributions:

\textbf{A multidimensional music aesthetic annotation framework.}  
We define an evaluation framework with one overall score and 10 fine-grained perceptual dimensions, supported by annotation guidelines to ensure consistency.

\textbf{A large-scale music aesthetics dataset and benchmark.}  
We construct MADB, a dataset of 9999 tracks annotated by trained annotators, each rated by 9–11 annotators across all dimensions, with additional textual comments.

\textbf{A unified benchmark for music aesthetic assessment.}  
We evaluate multiple pretrained models, including CLAP-based approaches, under a unified framework. Results show that current models capture only partial aesthetic information, revealing the difficulty of the task and the need for improved representations.

\section{Related Work}

\subsection{Music Aesthetic Evaluation and Datasets}

Early studies on music aesthetics in the MIR community mainly focused on perceptual modeling with handcrafted musical features. For instance, Marias et al. \cite{} adopted musical tension dynamics as a surrogate indicator of aesthetic quality. Subsequent work by Xin Jin et al. \cite{aesaudio,aesperf} further combined traditional feature engineering with neural networks to regress aesthetic scores from hand-designed acoustic features. Nevertheless, these methods are constrained by conventional MIR paradigms and exhibit limited representation capacity compared to modern Transformer-based models.

In contrast, aesthetic evaluation in image and video domains has achieved remarkable progress via Transformer architectures trained on large-scale expert-annotated corpora. Xin Jin’s team has built comprehensive aesthetic models for visual content \cite{paints, vadb, vga} and constructed large-scale datasets annotated by professionals from art and film institutions, enabling the learning of high-level aesthetic representations.

In the music domain, several benchmark datasets have been proposed, such as MusicEval \cite{musiceval} and SongEval \cite{songeval}, which provide basic benchmarks for music aesthetic evaluation. However, these datasets suffer from limited data scale, insufficient annotator participation, and incomplete coverage of perceptual dimensions. Most existing datasets only provide single-value aesthetic ratings and lack rich supervisory signals such as multi-dimensional perceptual scoring and textual subjective feedback, making them inadequate for capturing the complexity of human music aesthetics.

\subsection{Music Representation Learning}

Self-supervised learning has substantially advanced music representation learning in recent years \cite{cpc, wav2vec2}. Pre-trained models including MuQ \cite{muq} and MERT \cite{mert} learn universal audio representations from large-scale music corpora and yield promising transfer performance across downstream MIR tasks. Moreover, models such as PANNs \cite{panns} and HTSAT \cite{htsat} have also been widely adopted.
These pre-trained models can effectively capture diverse acoustic and musical characteristics, including timbre, rhythm, and structural patterns, which are inherently correlated with human aesthetic perception.

However, they are not explicitly optimized for perceptual or aesthetic modeling, nor do they incorporate structured aesthetic dimensions and subjective human evaluation signals. It remains an open problem to what extent general-purpose music representations can align with fine-grained human aesthetic judgments.

\subsection{Music-Text Cross-Modal Learning}

Multimodal research has made considerable progress in aligning music and textual representations. CLAP \cite{clap} learns unified audio-text embeddings via contrastive learning. As a foundational framework, CLIP \cite{clip} also verifies the efficacy of large-scale cross-modal alignment across perceptual domains. Extended frameworks such as CLAMP3 \cite{clamp3} further integrate structured textual attributes to enhance fine-grained cross-modal alignment.

Music-text multimodal models offer strong capability to associate audio content with semantic and contextual descriptions. Most existing methods primarily concentrate on coarse-grained attributes such as music genre and general mood. By contrast, rich textual subjective feedback, which is rarely exploited in current literature, holds great potential for modeling fine-grained perceptual and aesthetic traits, highlighting a promising direction for injecting human-centered perceptual knowledge into music understanding systems.

\section{MADB Dataset and Benchmark}
\label{headings}

\begin{table}[htbp]
\centering
\caption{Music Aesthetic Dimensions}
\label{tab:dimension_definition}

\begin{tabular}{|>{\raggedright\arraybackslash}p{0.12\linewidth}|>{\raggedright\arraybackslash}p{0.12\linewidth}|>{\raggedright\arraybackslash}p{0.66\linewidth}|}
\hline
\textbf{Type} & \textbf{Attribute} & \textbf{Interpretation} \\
\hline
\multirow{4}{=}{Composition Stage}& melody perception & Melodic perception refers to the pleasantness and structural coherence of a melody, including the naturalness of pitch progression and overall continuity.\\
\cline{2-3}
 & melody emotion &  Melodic emotion measures the intensity and expressiveness of emotional content conveyed by the melody, including clarity and emotional impact.\\
\cline{2-3}
 & rhythm perception &  Rhythmic perception refers to the appropriateness and clarity of rhythmic patterns, as well as their alignment with emotional progression.\\
\cline{2-3}
 & structure perception &  Structural perception refers to the clarity, diversity, and novelty of musical structure, as well as the contrast and organization between sections.\\
\hline
\multirow{2}{=}{Arrangement Stage}& arrangement perception &  Arrangement perception refers to the appropriateness and coherence of arrangement and instrumentation, including timbre selection, accompaniment, harmonic design, and overall sonic balance.\\
\cline{2-3}
 & arrangement emotion &  Arrangement emotion measures how well instrumentation and arrangement support emotional expression, including the alignment of timbre, orchestration, and atmosphere with the intended affect.\\
\hline
\multirow{3}{=}{Performance Stage}& performance and singing emotion &  Performance and singing emotion measures the clarity, intensity, and emotional impact conveyed through instrumental performance or singing.\\
\cline{2-3}
 & enunciation and singing skill &  Articulation and singing technique refers to lyrical clarity, vocal timbre, voice control, and the appropriateness and quality of singing techniques.\\
\cline{2-3}
 & performance skill &  Instrumental technique refers to the timbral quality and technical execution in instrumental performance, including techniques such as vibrato and tonguing.\\
\hline
Post-production Stage & sound effect perception &  Audio effects perception refers to the appropriateness of effects and mixing, including effects usage, track balance, and overall clarity and coherence of the sound.\\
\hline
Final production & overall score &  The overall score represents a holistic evaluation of the music’s aesthetic quality, reflecting the integrated perception of multiple attributes and the overall listening experience.\\
\hline
\end{tabular}
\end{table}

\subsection{Aesthetic Annotation Framework}

To model the multi-dimensional nature of music perception, we design a structured annotation framework consisting of 10 perceptual dimensions and one overall score. 
These dimensions capture complementary aspects of musical experience, including melody, arrangement, rhythm, structure, performance, and production-related attributes.

Not all dimensions are applicable to every track. For non-applicable cases (e.g., singing-related attributes for instrumental music), annotators assign a score of 0. 
These entries are retained as non-applicable labels and excluded from downstream model optimization.

The 10 dimensions cover key aspects of music production, including melody perception and emotion, rhythm and structure, arrangement quality and expressiveness, performance-related attributes such as emotion and technical skill, and post-production effects. These dimensions jointly reflect both low-level acoustic qualities and high-level perceptual judgments.

The dimensions are organized to align with the major stages of the music production pipeline, including composition, arrangement, performance, and post-production. 
This formulation provides a structured decomposition of music aesthetics and enables systematic analysis of different perceptual factors.

In addition to fine-grained annotations, we collect an overall aesthetic score representing the annotator’s holistic judgment. 
This score is not defined as a fixed aggregation of individual dimensions, but instead reflects a higher-level perceptual integration process.

\subsection{Semantic Annotations}

In addition to numerical ratings, MADB includes semantic annotations to enrich the representation of music aesthetics. Specifically, each track is associated with short textual comments and tag-based labels.

Annotators are required to provide brief textual comments(around 50 words) describing their aesthetic impressions of the music. These comments serve as complementary explanations for the numerical scores and provide insight into the reasoning process behind human aesthetic judgments. The comments were originally written in Chinese and subsequently translated into English using Qwen2.5-Instruct. Comments those less than 10 words will be ignored. 

We further define two tag-based annotation dimensions: genre and emotion. Each track is assigned up to two genre labels and up to two emotion labels from 10 genre labels and 6 emotion labels. These tags provide high-level semantic context and facilitate future research on multi-modal learning and cross-modal alignment. The complete list of tags is provided in the Appendix.

The combination of numerical ratings, textual comments, and semantic tags enables MADB to support both perceptual modeling and multi-modal aesthetic analysis.

\subsection{Data Sources}

The MADB dataset is constructed from four complementary sources to ensure diversity in musical style, production quality, and generation paradigm.

The first subset consists of 2,799 manually collected tracks curated from online platforms. This subset includes both professionally produced music (e.g., canonical works such as famous classical musics) and widely consumed popular content (e.g., short-video platform music), as well as tracks with varying production quality such as royalty-free or publicly available music. This design aims to capture a broad spectrum of real-world listening scenarios.

The second and third subsets consist of AI-generated music. Specifically, 1,000 tracks are generated by the Suno system, including a subset conditioned on existing song lyrics, and 1,800 tracks are generated by the Levo system. These samples introduce diverse generative styles and enable the evaluation of aesthetic perception in AI-generated content.

The fourth subset consists of 4,400 tracks from the Muchin dataset\cite{muchin}, which further expands coverage across musical genres and compositional styles.

This multi-source construction results in a dataset that spans both human-composed and AI-generated music, covering a wide range of production qualities, stylistic variations, and emotional characteristics. Such diversity is critical for establishing a robust and generalizable benchmark, enabling evaluation across different creation paradigms and varying aesthetic characteristics.

We provide all annotation data and a large subset(7030 songs)of publicly available audios, includes 2630 songs generated by AI, which can be downloaded on our huggingface page\footnote{\url{https://huggingface.co/datasets/sirui1/MADB-Dataset}}, and 4400 songs from open-source Muchin dataset, which can be downloaded on Muchin's github page\footnote{\url{https://github.com/CarlWangChina/MuChin}}. Other copyrighted audios are not allowed to be released due to licensing constraints. 

\subsection{Annotators}
The dataset is annotated by 30 annotators, each treated as an independent rating source in our analysis.

Each music track is independently evaluated by an average of 10 annotators, enabling robust estimation of inter-rater agreement and stable aggregation of aesthetic scores. This multi-annotator design helps reduce individual bias and supports reliable statistical analysis of perceptual consistency.

To ensure annotation quality, we adopt a three-stage quality control pipeline. The first stage consists of the primary annotation process. In the second stage, 20\% of the annotations are randomly sampled for review and consistency checking. In the third stage, an additional 10\% of the annotations are randomly selected for further validation by a different reviewer. The review and validation stages are conducted independently to reduce confirmation bias and improve reliability.

Annotators were recruited from multiple music conservatories, media universities, and industry partners across several institutions. All annotators had received at least three years of formal professional music training, and all held a bachelor’s degree or higher. This ensured that raters possessed sufficient musical knowledge and listening expertise to provide consistent and informed evaluations. Quality control and secondary review were conducted by senior experts with formal music training and professional evaluation experience, each with more than ten years of professional music education background.

\subsection{Annotation Statistics and Distribution}

\begin{figure}[htbp]
    \centering
    \includegraphics[width=1\linewidth]{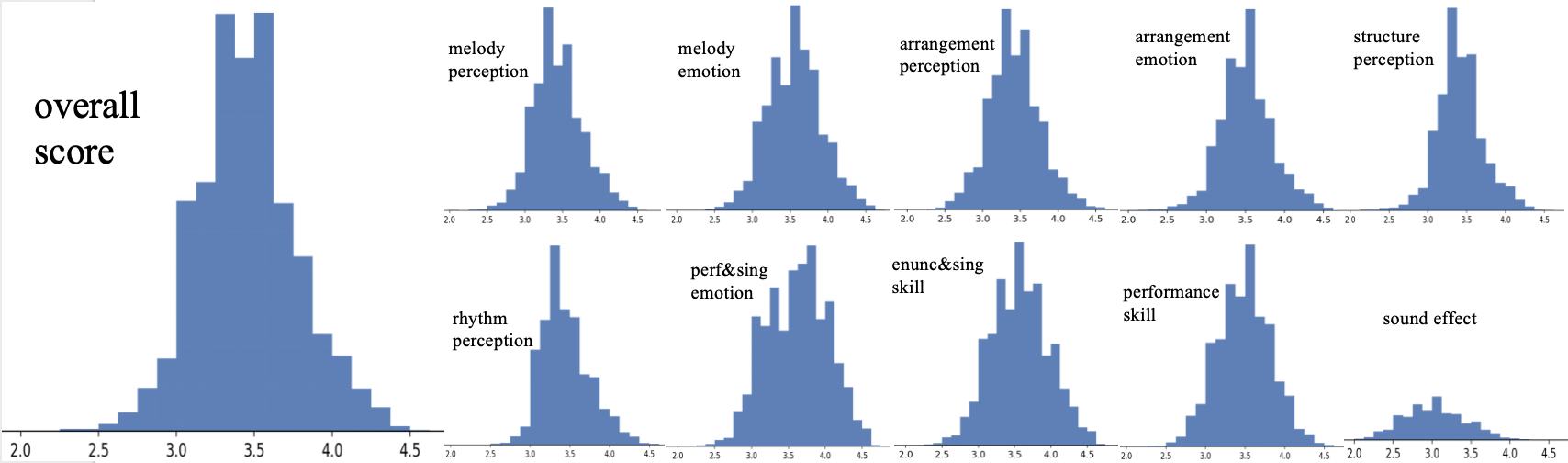}
    \caption{Histograms of average score across dimensions}
    \label{fig:hist}
\end{figure}

We analyze the distribution of aesthetic scores across all dimensions. 
As shown in Fig.~2, scores are predominantly concentrated in the mid-to-high range (approximately 2.5–4.5), with relatively few extreme values. 
This reflects a common characteristic of human aesthetic evaluation, where most music is perceived as moderately good rather than strongly polarized. 
Such a distribution implies a limited dynamic range, making fine-grained regression particularly challenging.

We further evaluate annotation consistency using the Intraclass Correlation Coefficient (ICCK) The results are all around 0.8,  indicate strong multi-rater reliability across most dimensions, supporting the stability and validity of the aggregated scores.

\section{Experiments}

\begin{figure}[htbp]
    \centering
    \includegraphics[width=1\linewidth]{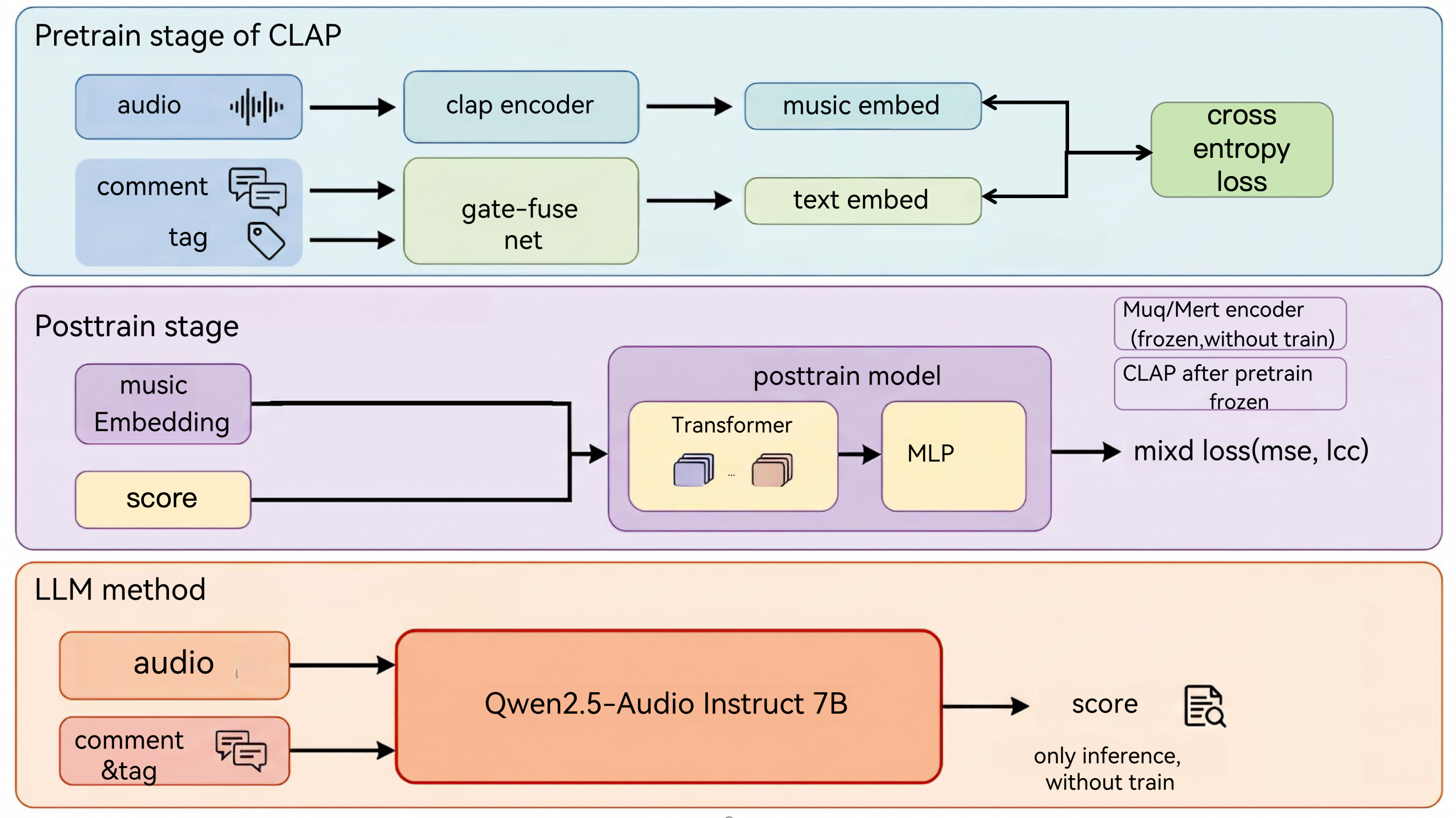}
    \caption{Three categories of evaluated methods: (1) pretrained encoders with post-training, (2) CLAP with semantic adaptation, and (3) LLM-based evaluation.}
    \label{fig:met}
\end{figure}

\subsection{Pretraining and Evaluation Framework}
We build upon the CLAP framework~\cite{clap} by incorporating semantic supervision from both textual comments and structured tags (e.g., genre and mood).

All comments are originally written in Chinese and translated into English using Qwen2.5-Instruct~\cite{qwen2.5} for CLAP's training, both original Chinese comments and translated English comments are provided. For each audio sample, multiple annotator comments are encoded and aggregated via mean pooling to form a unified textual representation. In parallel, genre and mood tags are embedded into low-dimensional vectors and projected into the same embedding space. The two representations are fused through a learnable gating mechanism to produce a joint semantic embedding.

To enhance audio–semantic alignment, we perform a pre-adaptation stage based on contrastive learning. Specifically, the audio encoder is trained to align with the fused semantic representation using a cosine similarity-based objective. This stage relies only on comments and tags, without using aesthetic scores, thereby avoiding label leakage while encouraging the model to capture perceptual and semantic information.

For downstream evaluation, we adopt a unified regression framework across all models. Audio representations extracted from pretrained encoders are temporally pooled into fixed-length embeddings, followed by a Transformer-based regression head to predict aesthetic scores. Models are trained using mean squared error (MSE) as the primary objective, optionally combined with a correlation-based loss to better align predictions with human perceptual rankings.

We evaluate multiple encoder variants, including MuQ, MERT, and CLAP, as well as CLAP adapted with comment-only and comment–tag supervision. All models share the same downstream architecture, differing only in encoder and pretraining strategy, enabling a controlled comparison of representation quality. The encoders are frozen during downstream training. 

We randomly split the dataset into training and validation sets with a fixed ratio. All models are optimized using Adam with a consistent set of hyperparameters across experiments. During training, zero values indicate non-applicable dimensions and are excluded from optimization, with separate models trained when necessary. All clap-based experiments are conducted with 4 fixed random seeds(42,120,5,2500) to ensure reproducibility and to reduce the influence of stochastic factors such as data shuffling, parameter initialization, and mini-batch sampling. Muq and Mert experiments are under seed 42.

All experiments are conducted on NVIDIA A100 GPUs for embedding extraction and pretraining, and NVIDIA Tesla V100 GPUs for downstream training.

\begin{table*}[htbp]
\caption{Performance comparison across different metrics and aesthetic dimensions.(mean+std) \textbf{}}
\label{tab:main_results}
\centering

\normalsize

\setlength{\tabcolsep}{2pt} 
\renewcommand{\arraystretch}{1.5} 

\resizebox{\textwidth}{!}{
\begin{tabular}{llccccccccccc}
\toprule
Metric & Method 
& Overall\_Score
& Mel.\ Perc. 
& Mel.\ Emo. 
& Arr.\ Perc. 
& Arr.\ Emo. 
& Rhy.\ Perc.
& Struc.\ Perc.
& Perf\_Sing.\ Emo. 
& Sing.\ Skill 
& Perf.\ Skill 
& Sound Effect \\
\midrule

\multirow{5}{*}{MSE$\downarrow$}
& MERT           & 1& 0.090 & 0.083 & 0.111 & 0.108 & 0.079 & 0.098 & 0.079 & 0.096 & 0.091 & 0.135 \\
& MuQ            & \textbf{0.081} & \textbf{0.070} & \textbf{0.075} & \textbf{0.064} & \textbf{0.062} & \textbf{0.063} & \textbf{0.081} & \textbf{0.076} & \textbf{0.075} & \textbf{0.074} & \textbf{0.097} \\
\cdashline{2-13}
& CLAP           & 0.109±0.001& 0.110±0.005& 0.121±0.009& 0.125±0.006& 0.123±0.010& 0.103±0.005& 0.101±0.008& 0.135±0.004& 0.145±0.006& 0.121±0.004& 0.158±0.007\\
& CLAP+C       & \underline{0.108±0.007}& 0.111±0.005& \underline{0.118±0.003}& 0.123±0.008& 0.125±0.008& 0.105±0.003& 0.100±0.004& \underline{0.134±0.005}& \underline{0.139±0.002}& 0.123±0.002& \underline{0.153±0.006}\\
& CLAP+C\&T  & 0.109±0.004& \underline{0.109±0.006}& 0.120±0.009& \underline{0.120±0.008}& \underline{0.123±0.003}& \underline{0.097±0.003}& \underline{0.098±0.003}& 0.134±0.008& 0.144±0.003& \underline{0.121±0.005}& 0.158±0.005\\
\midrule

\multirow{5}{*}{LCC$\uparrow$}
& MERT           & 0.626 & 0.670 & 0.713 & 0.561 & 0.567 & 0.674 & 0.534 & 0.782 & 0.674 & 0.635 & 0.600 \\
& MuQ            & \textbf{0.718} & \textbf{0.715} & \textbf{0.766} & \textbf{0.680} & \textbf{0.685} & \textbf{0.727} & \textbf{0.650} & \textbf{0.820} &\textbf{ 0.748} & \textbf{0.713} & \textbf{0.661} \\
\cdashline{2-13}
& CLAP           & 0.436±0.007& 0.443±0.010& 0.464±0.005& 0.335±0.009& 0.345±0.007& 0.419±0.027& 0.327±0.010& 0.576±0.003& 0.488±0.006& 0.411±0.009& 0.463±0.018\\
& CLAP+C       & 0.428±0.011& 0.442±0.008& 0.459±0.008& 0.350±0.020& 0.343±0.012& 0.426±0.017& 0.334±0.010& \underline{0.577±0.007}& 0.483±0.010& 0.417±0.011& 0.463±0.014\\
& CLAP+C\&T  & \underline{0.445±0.014}& \underline{0.448±0.012}& \underline{0.478±0.004}& \underline{0.355±0.023}& \underline{0.349±0.014}& \underline{0.426±0.008}& \underline{0.347±0.013}& 0.576±0.009& \underline{0.492±0.006}& \underline{0.427±0.020}& \underline{0.475±0.012}\\
\midrule

\multirow{5}{*}{SRCC$\uparrow$}
& MERT           & 0.626 & 0.676 & 0.726 & 0.567 & 0.575 & 0.662 & 0.509 & 0.790 & 0.665 & 0.643 & 0.613 \\
& MuQ            & \textbf{0.714} & \textbf{0.716} & \textbf{0.775} & \textbf{0.668} & \textbf{0.682} & \textbf{0.655} & \textbf{0.610} & \textbf{0.822} & \textbf{0.740} & \textbf{0.707} & \textbf{0.668} \\
\cdashline{2-13}
& CLAP           & 0.379±0.008& 0.394±0.021& 0.427±0.015& 0.335±0.011& 0.322±0.010& 0.332±0.028& 0.270±0.012& 0.487±0.010& 0.410±0.014& 0.394±0.011& 0.476±0.015\\
& CLAP+C       & 0.372±0.025& 0.394±0.017& 0.414±0.003& 0.344±0.018& \underline{0.325±0.018}& \underline{0.342±0.017}& 0.282±0.016& 0.493±0.009& 0.400±0.015& 0.391±0.014& 0.473±0.008\\
& CLAP+C\&T  & \underline{0.388±0.018}& \underline{0.405±0.013}& \underline{0.443±0.017}& \underline{0.350±0.025}& 0.325±0.025& 0.342±0.022& \underline{0.295±0.015}& \underline{0.502±0.019}& \underline{0.420±0.010}& \underline{0.405±0.020}& \underline{0.482±0.003}\\
\midrule

\multirow{5}{*}{KRCC$\uparrow$}
& MERT           & 0.451 & 0.490 & 0.533 & 0.404 & 0.407 & \textbf{0.484} & 0.356 & 0.588 & 0.483 & 0.464 & 0.447 \\
& MuQ            & \textbf{0.528} & \textbf{0.525} & \textbf{0.584} & \textbf{0.492} & \textbf{0.501} & 0.480 & \textbf{0.443} & \textbf{0.628} & \textbf{0.550} & \textbf{0.521} & \textbf{0.478} \\
\cdashline{2-13}
& CLAP           & 0.259±0.007& 0.271±0.014& 0.301±0.020& 0.224±0.007& 0.220±0.007& 0.229±0.028& 0.185±0.009& 0.339±0.010& 0.282±0.011& 0.268±0.009& 0.327±0.011\\
& CLAP+C       & 0.255±0.018& 0.271±0.013& 0.285±0.001& 0.231±0.011& \underline{0.222±0.012}& \underline{0.236±0.019}& 0.193±0.011& 0.344±0.007& 0.275±0.011& 0.267±0.010& 0.322±0.006\\
& CLAP+C\&T  & \underline{0.266±0.012}& \underline{0.279±0.010}& \underline{0.312±0.024}& \underline{0.235±0.017}& 0.221±0.017& 0.235±0.021& \underline{0.202±0.012}& \underline{0.349±0.016}& \underline{0.289±0.006}& \underline{0.277±0.015}& \underline{0.330±0.004}\\
\bottomrule
\end{tabular}
}

\end{table*}















\subsection{Result Performance}

Table~\ref{tab:main_results} presents the performance of different models across 11 aesthetic dimensions under four evaluation metrics with mean and std.  \textbf{Bold} indicates the best result among all methods, and \underline{underlined} values denote the best performance within CLAP-based methods.

Overall, models based on MuQ and MERT significantly outperform CLAP-based variants across all metrics, achieving substantially lower MSE and higher correlation scores (LCC, SRCC, and KRCC). 
In particular, MuQ consistently achieves the best performance on most dimensions, indicating that current audio representation models specifically designed for music understanding are more suitable for aesthetic prediction tasks.

In contrast, CLAP-based models achieve comparatively lower performance, particularly on perceptual and structural dimensions such as arrangement and structure perception.

Despite the use of strong pretrained encoders, the overall performance remains far from saturation, indicating that multi-dimensional aesthetic prediction is a challenging task.

\subsection{Effect of Comment and Tag-Based Adaptation}

We further compare three CLAP-based\cite{clap} variants: the original pretrained CLAP, CLAP adapted using textual comments (CLAP+Com), and CLAP adapted using both comments and tag fusion (CLAP+Tag\&Com). 
Across most dimensions and metrics, both adapted variants outperform the original CLAP model, as indicated by consistently better MSE and higher correlation values.

This improvement demonstrates that incorporating human-provided semantic signals—especially free-form comments—provides useful supervisory information for aesthetic prediction. 
Compared to comment-only adaptation, the addition of structured tag information further improves performance on several dimensions (e.g., melody and rhythm perception), indicating that tags offer complementary and more stable semantic priors.

Importantly, this adaptation is performed without using any aesthetic scores, ensuring that the observed improvements are not due to label leakage but rather to enhanced representation learning. 
These findings confirm the value of combining textual and structured annotations, and suggest that multi-source semantic supervision is a promising direction for improving music aesthetic modeling.

\subsection{Multimodal LLM-Based Aesthetic Prediction}

To analyze the contribution of different modalities to music aesthetic evaluation, we conduct a zero-shot probing experiment using \textit{Qwen2-Audio-7B-Instruct}. We consider three input configurations: \textbf{Audio only}, \textbf{Comment\&Tag}, and \textbf{Audio + Comment\&Tag}.

For each sample, the model predicts an overall aesthetic score (1–5) based on an instruction prompt aligned with the perceptual dimension definitions in Table~\ref{tab:dimension_definition}. To ensure fair comparison, each configuration uses a separate input pipeline, explicitly controlling which modalities are provided without describing missing inputs.

\begin{table*}[htbp]
\caption{Performance comparison across different metrics and aesthetic dimensions of LLM.}
\label{tab:llm_results}
\centering

\normalsize
\setlength{\tabcolsep}{4pt} 
\renewcommand{\arraystretch}{1.4} 

\resizebox{\textwidth}{!}{
\begin{tabular}{llccccccc}
\toprule
Metric & Method 
& Overall
& Mel.\ Perc. 
& Mel.\ Emo. 
& Rhy.\ Perc.
& Struc.\ Perc.
& Perf.\ Emo. 
& Perf.\ Skill \\
\midrule

\multirow{3}{*}{MSE$\downarrow$}
& Audio only           & 1.181 & 1.206 & 1.060 & 1.249 & 1.007    & 1.049     & 1.214 \\
& Comment\&Tag         & \textbf{0.091} & \textbf{0.112} & \textbf{0.094} & \textbf{0.149} & \textbf{0.118} & \textbf{0.109} & \textbf{0.123} \\
& Audio+Comment\&Tag   & 0.095 & 0.113 & 0.101 & 0.154 & 0.119 & 0.111 & 0.133\\
\midrule

\multirow{3}{*}{LCC$\uparrow$}
& Audio only           & -0.001 & 0.003 & -0.038 & -0.029 & -0.005    & -0.036     & -0.060 \\
& Comment\&Tag         & \textbf{0.736} & 0.685 & 0.705 & 0.528 & \textbf{0.667} & 0.660 & 0.635 \\
& Audio+Comment\&Tag   & 0.734 & \textbf{0.694} & \textbf{0.707} & \textbf{0.541} & 0.649 & \textbf{0.698} & \textbf{0.660}\\
\midrule

\multirow{3}{*}{SRCC$\uparrow$}
& Audio only           & -0.035 & -0.040 & -0.078 & -0.044 & -0.027    & -0.111     & -0.098 \\
& Comment\&Tag         & 0.717 & 0.669 & 0.690 & 0.508 & \textbf{0.623} & 0.662 & 0.635 \\
& Audio+Comment\&Tag   & \textbf{0.718} & \textbf{0.686} & \textbf{0.701} & \textbf{0.514} & 0.610 & \textbf{0.670} & \textbf{0.650}\\
\midrule

\multirow{3}{*}{KRCC$\uparrow$}
& Audio only           & -0.028 & -0.032 & -0.063 & -0.036 & -0.022    & -0.088     & -0.080 \\
& Comment\&Tag         & 0.593 & 0.552 & 0.569 & 0.408 & \textbf{0.513} & 0.541 & 0.516 \\
& Audio+Comment\&Tag   & \textbf{0.594} & \textbf{0.561} & \textbf{0.573} & \textbf{0.409} & 0.496 &\textbf{ 0.568} & \textbf{0.522}\\
\bottomrule
\end{tabular}
}

\end{table*}

\subsection{Analysis of Multimodal Input Configurations}

As shown in Table~\ref{tab:main_results}, the \textbf{Audio+Comment\&Tag} and \textbf{Comment\&Tag }setting significantly outperforms \textbf{Audio-only}, achieving strong correlation. This indicates that MADB's human textual feedback provides highly informative signals for aesthetic judgment. This is because comments already encode high-level perceptual judgments and evaluative semantics produced by human annotators, effectively serving as a compressed representation of aesthetic reasoning that is closely aligned with the target scores. English comments are used in these experiments, which prove the reliability of translations.

Combining modalities (\textbf{Audio + Comment\&Tag}) yields further improvements on several metrics, suggesting that audio provides complementary information beyond textual descriptions. However, the relatively small gains also highlight the difficulty of extracting high-level aesthetic judgments directly from raw audio.

Overall, these results show that (1) human comments serve as a strong supervision signal, and (2) multimodal inputs lead to the most consistent performance. Notably, all results are obtained in a zero-shot setting, reflecting intrinsic modality informativeness rather than dataset-specific training.

\subsection{Discussion}

The gap between CLAP-based models and music-specific encoders (MuQ, MERT) suggests that aesthetic evaluation requires not only semantic alignment but also strong intrinsic audio representation capabilities.

 At the same time, the consistent gains from comment and tag-based adaptation, as well as the complementary signals from LLM-derived text representations, indicate that human annotations provide meaningful information that can partially bridge this gap.

Overall, the results validate the proposed benchmark as both challenging and informative, revealing clear limitations of existing models while highlighting the potential of leveraging richer annotation sources and lightweight language modeling for future improvements.

\section{Limitations}

Despite its scale and richness, MADB has several limitations. First, the dataset exhibits an imbalanced distribution of music styles, with a predominance of popular genres. While this reflects real-world listening patterns, it may bias models and limit generalization to underrepresented styles.

Second, textual comments are originally written in Chinese and translated into English for compatibility with pretrained language models, which may introduce noise or subtle semantic shifts.

Third, annotations are provided at the whole-track level without temporal granularity. In practice, aesthetic quality often varies across different segments within a piece, and the absence of segment-level annotations limits the ability to model such dynamic perceptual variations. Incorporating fine-grained temporal annotations is therefore an important direction for future work.

Fourth, the current tag system is relatively coarse and does not capture fine-grained stylistic distinctions (e.g., subgenres such as house or post-rock). Expanding the tag taxonomy with more detailed and hierarchical categories is an important direction for improving semantic expressiveness.

Finally, the dataset does not explicitly capture causal relationships between perceptual dimensions and overall aesthetic judgments, leaving this as an open problem for future research.

\section{Conclusion}

We study music aesthetic assessment, an increasingly important problem in the era of generative music.
Our analysis shows that it is inherently challenging due to its multi-dimensional, subjective, and temporally dependent nature, with concentrated score distributions, moderate inter-annotator agreement, and complex relationships across perceptual dimensions.

To support this direction, we introduce MADB, a large-scale dataset and benchmark with multi-dimensional ratings, textual comments, and semantic tags.
Based on this benchmark, we show that while modern pretrained models capture partial aesthetic signals, a substantial gap to human perception remains.

These findings suggest that music aesthetic evaluation requires not only strong audio representations but also deeper integration of perceptual, semantic, and contextual information.
Textual comments provide complementary signals but are insufficient alone, highlighting the need for multimodal approaches.

We hope MADB will advance research in human-centered music understanding and improve both generative models and evaluation frameworks.
Future work may explore tighter integration with large language models and more fine-grained temporal annotations to better capture human aesthetic reasoning.
It also opens opportunities for studying structured aesthetic reasoning and controllable generation grounded in human perception.
While offering benefits for recommendation and creative tools, such systems may introduce bias and should be used responsibly.
\newpage
\bibliographystyle{unsrtnat}
\bibliography{reference}


\appendix
\newpage
\section{Technical appendices and supplementary material}

\subsection{Inter-Annotator Agreement Across Dimensions}

Figure~4 presents the ICCK values across all perceptual dimensions. Overall, the agreement is consistently high, with most dimensions exceeding 0.80, indicating strong multi-rater reliability.

Structurally grounded dimensions, such as \textit{melody\_emotion} and \textit{enunciation\_and\_singing\_skill}, achieve relatively higher agreement (e.g., above 0.83), suggesting that annotators share consistent judgments on well-defined perceptual attributes. In contrast, more subjective or context-dependent dimensions, such as \textit{sound\_effect\_perception}, exhibit comparatively lower agreement (0.72), reflecting the inherent ambiguity in evaluating production-related effects.

Notably, the overall score also demonstrates high reliability (ICCK = 0.84), indicating that despite its holistic nature, annotators maintain consistent global judgments. These results support the stability and validity of the annotation process.

\begin{figure}[htbp]
    \centering
    \includegraphics[width=\linewidth]{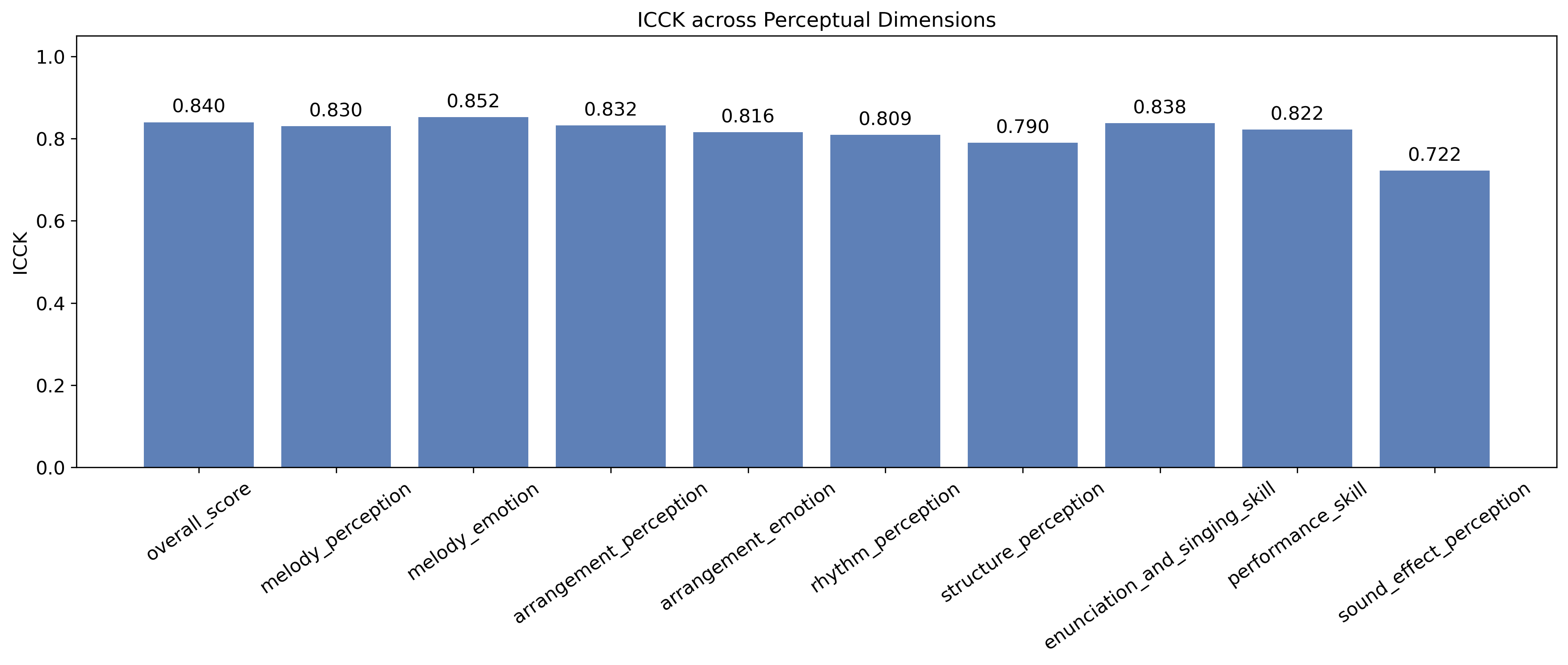}
    \caption{Inter-rater agreement measured by ICC$_k$ across dimensions.}
    \label{fig:icc}
\end{figure}

\subsection{Score Distribution Across Data Sources}

Figure~5 illustrates the mean overall scores across different data sources. We observe noticeable variation in average ratings: \textit{internet} and \textit{suno} samples achieve relatively higher scores (around 3.5), while \textit{levo} samples receive lower ratings (approximately 3.18). 

Despite these differences in mean values, all sources exhibit non-negligible standard deviations, indicating substantial intra-source diversity. This suggests that each source contains a mixture of high- and low-quality samples, rather than being dominated by uniformly strong or weak content.

These findings highlight that while source-level biases exist, the dataset maintains sufficient variability within each source, contributing to the overall complexity and realism of the benchmark.

\begin{figure}[htbp]
    \centering
    \includegraphics[width=0.7\linewidth]{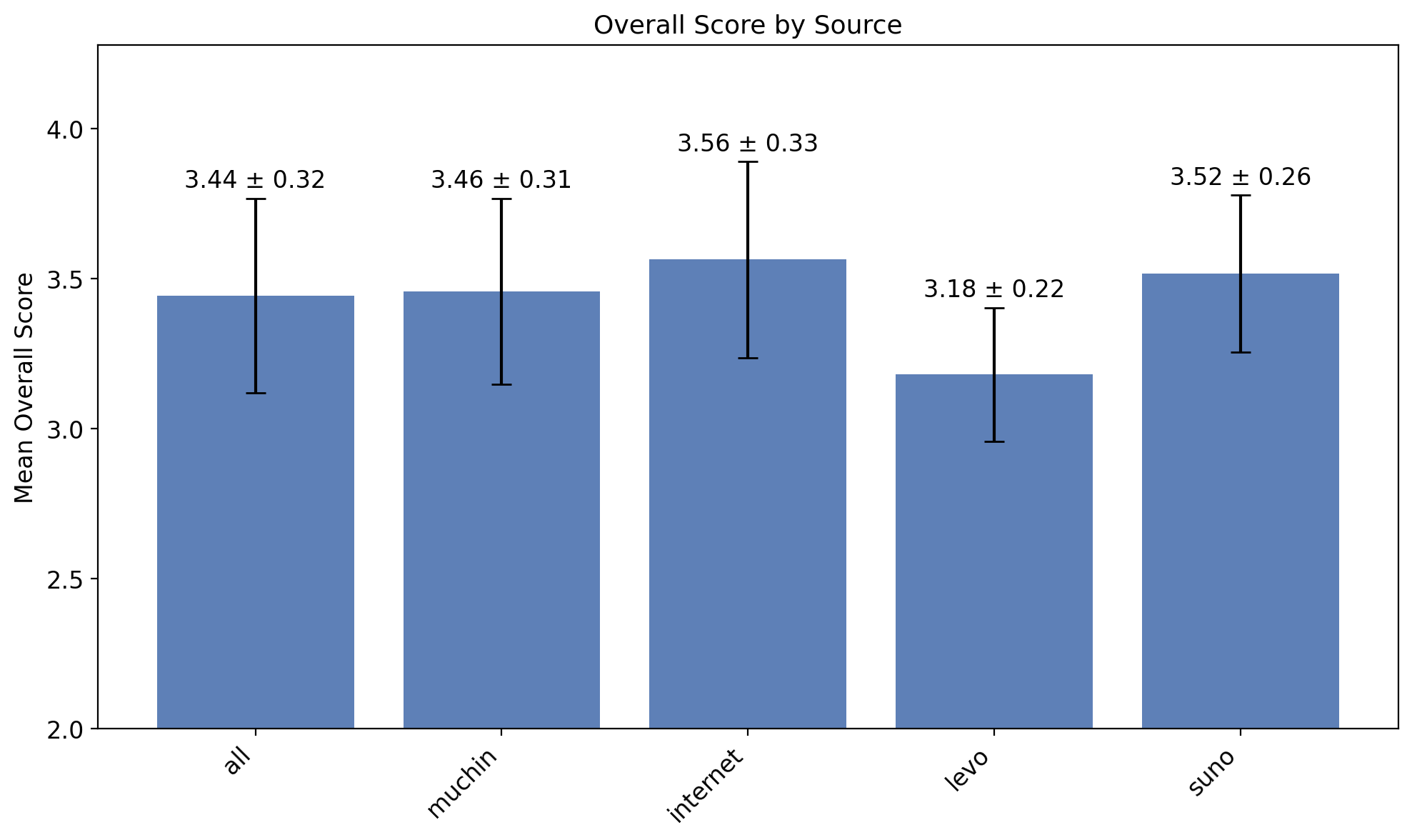}
    \caption{Average overall score and standard deviation across data sources.}
    \label{fig:avg_score}
\end{figure}

\subsection{Annotators Background}


\begin{table}[htbp]
\centering
\caption{Annotator Background Summary}
\label{tab:annotator_background}
\small
\begin{tabular}{lll}
\toprule
Category & Institutions & Count \\
\midrule

\multirow{4}{*}{Music Conservatories}
& Central Conservatory of Music, China &  \multirow{4}{*}{10} \\
& Tianjin Conservatory of Music &  \\
& Shenyang Conservatory of Music &  \\
& Sichuan Conservatory of Music &  \\

\midrule

\multirow{3}{*}{Media Universities}
& Communication University of China & \multirow{3}{*}{4} \\
& School of Journalism and Communication, Chongqing Normal University &  \\
& Communication University of Zhejiang &  \\

\midrule

\multirow{3}{*}{Comprehensive Universities}
& The School of Music, CUHK-Shenzhen &  \multirow{3}{*}{5} \\
& College of Arts and Media, Tongji University &  \\
& School of Music, Jiangxi Science and Technology Normal University &  \\

\midrule

\multirow{1}{*}{Industry Practitioners}
& Music Industry Practitioners & \multirow{1}{*}{11}  \\

\bottomrule
\end{tabular}
\end{table}

\subsection{Tags Definition}

\begin{table}[htbp]
\centering
\caption{Emotion categories used in MADB.}
\label{tab:emotion_categories}
\begin{tabular}{p{0.22\linewidth}p{0.70\linewidth}}
\toprule
Emotion & Description \\
\midrule

Happy & Music conveying positive, joyful, and uplifting emotions. This category often includes cheerful melodies, bright harmonic colors, energetic rhythms, and a sense of celebration, excitement, or pleasure. \\
\midrule
Peaceful & Music associated with calmness, serenity, warmth, and relaxation. It is often characterized by gentle dynamics, smooth melodic motion, soft timbres, and a stable or spacious atmosphere. \\
\midrule
Passionable & Music expressing passion, intensity, excitement, and grandeur. This category includes energetic, heroic, or highly motivating emotional qualities, often supported by strong rhythmic drive, dynamic contrast, and powerful performance. \\
\midrule
Sad & Music conveying sorrow, melancholy, loneliness, or reflective emotional states. It often features slower tempos, minor tonalities or darker harmonic colors, expressive phrasing, and introspective atmosphere. \\
\midrule
Angry & Music expressing anger, aggression, frustration, or emotional release. It may involve harsh timbres, strong accents, intense dynamics, distorted textures, or highly forceful rhythmic patterns. \\
\midrule
Nervous & Music associated with tension, anxiety, suspense, or unease. This category may include unstable harmony, irregular rhythm, sharp contrasts, dissonant textures, or eerie sonic atmospheres. \\

\bottomrule
\end{tabular}
\end{table}

\begin{table}[htbp]
\centering
\caption{Genre categories used in MADB.}
\label{tab:genre_categories}
\begin{tabular}{p{0.22\linewidth}p{0.70\linewidth}}
\toprule
Genre & Description \\
\midrule
Pop & Mainstream popular music with accessible melodies, clear verse-chorus structures, and polished production. This category  typically emphasizing catchy hooks, stable rhythmic grooves, and broad audience appeal. \\
\midrule
Chinese Pop & Chinese-language popular music that combines mainstream pop songwriting with elements of Chinese musical tradition. It may incorporate pentatonic melodic patterns, Chinese instruments or timbral references, distinctive vocal delivery, and harmonic or arrangement choices different from Western classic pop. \\
\midrule
Country & A popular music style rooted in American folk and rural traditions, commonly featuring acoustic guitar, steady rhythmic patterns, narrative lyrics, and themes such as daily life, nostalgia, love, and personal experience. \\
\midrule
Rock/Metal & Guitar-centered music characterized by strong rhythmic drive, amplified instrumentation, and expressive vocal or instrumental performance. This category also includes heavier substyles such as metal, which often feature distorted guitars, dense textures, and higher intensity. \\
\midrule
Electronic & Music primarily produced or shaped through electronic sound synthesis, sampling, sequencing, and digital production. It often emphasizes timbre design, texture, atmosphere, and programmed rhythmic patterns. \\
\midrule
DJ & Dance-oriented electronic music designed for clubs, festivals, or continuous mixing contexts. Compared with general electronic music, this category places stronger emphasis on beat-driven structures, drops, build-ups, repeated grooves, and high-energy rhythmic progression. \\
\midrule
Classical & Western classical music, including orchestral works, chamber music, solo instrumental pieces, and vocal compositions. It typically emphasizes formal structure, thematic development, instrumental technique, and long-range musical organization. \\
\midrule
Chinese Classical & Traditional Chinese or Chinese-style classical music, including works based on Chinese instruments, traditional modes, pentatonic materials, regional idioms, and culturally specific expressive techniques. It differs from Chinese Pop by placing greater emphasis on traditional musical language rather than popular-song structure. \\
\midrule
Blues/Jazz/Souls & Music derived from jazz, blues, and soul traditions, often characterized by swing or groove-based rhythm, extended harmony, improvisation, expressive phrasing, and rich timbral nuance. \\
\midrule
Hiphop/Rap & Vocal-centered music based on rhythmic speech, flow, rhyme, and beat-driven accompaniment. This category includes hip-hop and related rap styles, where lyrical delivery, rhythmic articulation, and production groove are central aesthetic features. \\
\bottomrule
\end{tabular}
\end{table}

\end{document}